\documentclass[aps,prl,reprint]{revtex4-1}
\usepackage{blindtext}
\usepackage{graphicx}% Include figure files
\usepackage{dcolumn}% Align table columns on decimal point
\usepackage{bm}% bold math
\usepackage{color}
\usepackage[T1]{fontenc}% 
\usepackage{mathrsfs,amsfonts,dsfont}
\usepackage{amstext}
\usepackage{mathtools}
\usepackage{enumitem}
\graphicspath{{graphics/}}
\usepackage{rotating} % for diagram- descriptions
\usepackage[english]{babel}  

\usepackage{wrapfig}
\usepackage{lipsum} 

\usepackage{braket}
\usepackage{bbm} 
\usepackage[colorlinks=true,citecolor=blue,linkcolor=blue]
{hyperref}

\begin{document}
\title{Activation of nonlocality in bound entanglement}
\author{Lucas Tendick}
\email{Lucas.Tendick@hhu.de}
\author{Hermann Kampermann }
\author{Dagmar Bru\ss}
\affiliation{Institute for Theoretical Physics III, Heinrich-Heine-Universit\"at D\"usseldorf, D-40225 D\"usseldorf, Germany}

\begin{abstract}
We discuss the relation between entanglement and nonlocality in the
hidden nonlocality scenario.
Hidden nonlocality signifies nonlocality that can be activated by applying
local filters to a particular state that admits a local hidden-variable
model in the Bell scenario.
We present a fully-biseparable three-qubit bound entangled state
with a local model for the most general (non-sequential) measurements. This proves for the first time that bound entangled states can admit a local model for general measurements.
We furthermore show that the local model breaks down when suitable local filters are applied.
Our results demonstrate the first example of activation of nonlocality in bound entanglement.
Hence, we show that genuine hidden nonlocality does not imply entanglement distillability.
\end{abstract}

\maketitle

\begin{figure}[h!]
   \includegraphics[scale = 0.43]{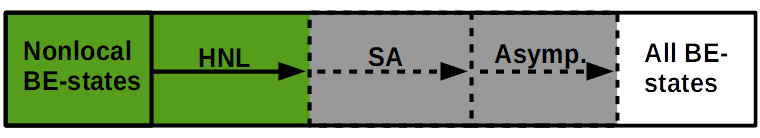} 
   \caption{Abstract overview of our results. We show that the set of nonlocal bound entangled states (BE-states) can be enlarged in the hidden nonlocality scenario (HNL). This is the first step towards a possible equivalence of all BE-states and all nonlocal BE-states. Further enlargements of the set of nonlocal BE-states could be provided by superactivation (SA) and the asymptotic scenario (Asymp.), similar to the case for distillable states. It is also an open question, whether the set can be enlarged to all BE-states in such scenarios. }
   \label{Results_sets}
 \end{figure}

Performing local measurements on certain entangled quantum states can lead to the phenomenon of quantum nonlocality. That is, the correlations obtained from the measurements are not compatible 
with the principle of local realism, witnessed by the violation of a so-called Bell inequality \cite{Bell_original}. Although entanglement and nonlocality were extensively studied since the foundation of quantum theory \cite{Bell_nonlocality,RevModPhys.81.865}, the relation between both is still not fully understood. \\
\indent After the seminal work by Bell \cite{Bell_original} as answer to the EPR-Gedankenexperiment \cite{PhysRev.47.777}, it was widely believed that entanglement and nonlocality are just two different notions of the inseparability of quantum states. Indeed, for pure entangled states nonlocality is a generic feature
\cite{Gisin_purestate_nonlocality,POPESCU1992293}. However, Werner \cite{PhysRevA.40.4277} showed that there exist mixed entangled states (so-called Werner states) which admit a local model for projective measurements. Later, Barrett \cite{PhysRevA.65.042302} extended this result by showing that certain 
Werner states admit a local model even when positive-operator valued measures (POVMs), i.e. most general non-sequential measurements are considered. This displays the inequivalence of entanglement and nonlocality in the Bell scenario. \\
\indent It was first noticed by Popescu \cite{PhysRevLett.74.2619} and more recently by Hirsch et al. \cite{PhysRevLett.111.160402} that some local entangled states can violate a Bell inequality when the observers apply judicious local filters as probabilistic pre-selection before the Bell test.
This phenomenon is referred to as hidden nonlocality, or as genuine hidden nonlocality when one considers an entangled quantum state $ \rho $ with a local model even for POVMs. However, it was shown that genuine hidden nonlocality is no general feature \cite{1367-2630-18-11-113019}. For example, a particular two-qubit Werner state remains local even after arbitrary local filtering. \\
\indent Note that hidden nonlocality is not the only extension of the Bell scenario. For instance, nonlocality can also be superactivated \cite{PhysRevA.87.042104} by allowing the parties to perform joint measurements on multiple copies of a local entangled state. An even more general concept is that of entanglement distillation \cite{RevModPhys.81.865}. In this scenario the parties have access to both, local filters and multiple copies of a given state, with the goal to probabilistically obtain pure entangled states. Distillable states can therefore always be seen as nonlocal resource in the so-called asymptotic scenario \cite{PhysRevLett.97.050503}. However, there exist entangled states which are not distillable to pure entangled states. These states build the famous set of bound entangled states \cite{PhysRevLett805239}. Studying the nonlocal properties of bound entangled states will approach the answer of the fundamental open question of whether all entangled states are nonlocal resources. Since bound entanglement is the weakest form of entanglement, it was conjectured by Peres \cite{quant-ph/9807017} that bound entangled states cannot lead to any nonlocal correlations at all. However, nowadays we know that the Peres conjecture is false \cite{1405.4502,PhysRevLett.108.030403}: bound entangled states can violate a Bell inequality. Despite these results and more advanced scenarios \cite{PhysRevA.86.052115}, about the activation of local bound entanglement nothing is known. \\
\indent In this letter, we answer the open question of whether bound entangled states with genuine hidden nonlocality exist in the affirmative. Specifically, we show that a certain three-qubit bound entangled state with a local model for POVMs can violate a Bell inequality when local filters were applied. This proves that genuine hidden quantum nonlocality does not imply entanglement distillability. Our results and possible extensions are visualized in Fig. \ref{Results_sets}.
 \newpage

\indent \textit{Preliminaries.}\textemdash Consider three distant parties Alice, Bob, and Charlie sharing an entangled quantum state $ \rho $. 
The parties can perform local measurements via the positive semidefinite operators $ M_{a \vert x}, M_{b \vert y}$, and $ M_{c \vert z} $ with the settings $ x, y, z $ and the outcomes $ a, b, c $. These operators form POVMs, as they satisfy the completeness relation $ \sum_a M_{a \vert x} = \mathds{1} $ (and analogously for Bob and Charlie), where $ \mathds{1} $ denotes the identity operator.
The resulting statistics is given by
\begin{align}
p(abc \vert xyz) = \mathrm{Tr}[(M_{a \vert x} \otimes M_{b \vert y} \otimes M_{c \vert z}) \rho]. \label{distribution}
\end{align}

The state $ \rho $ is said to be local (for $ \lbrace M_{a \vert x} \rbrace,\lbrace M_{b \vert y} \rbrace  $, and $ \lbrace M_{c \vert z} \rbrace $) if the distribution (\ref{distribution})
admits a local decomposition of the following form:

\begin{align}
 p(abc \vert xyz) = \int \pi (\lambda) p(a \vert x \lambda) p(b \vert y \lambda) p(c \vert z \lambda)  d \lambda.
 \label{LHV}
\end{align}

That is, the statistics can be explained by a local hidden-variable model (LHV), where $ \lambda \in \mathbb{R}$ is the shared local hidden-variable, distributed according to the density $ \pi (\lambda) $ such that $  \int \pi (\lambda) d \lambda = 1 $. The probability distributions $ p(a \vert x \lambda)$, $ p(b \vert y \lambda)$, and $ p(c \vert z \lambda) $ 
are typically called local response functions in this context. A state $ \rho $ with such a decomposition \emph{for all} possible measurements cannot violate any Bell inequality; otherwise it does violate
(at least) one Bell inequality.  \\
\indent A concept which is easier to handle and necessary for Bell nonlocality is the concept of quantum steering \cite{0034-4885-80-2-024001}. The steering scenario is an asymmetric scenario where one or more parties remotely steer the state of the remaining parties by performing measurements on their part of the state. Here, we focus on the so-called one-sided steering scenario where Alice tries to steer Bob and Charlie. We say a state does demonstrate steering if its probability distribution does not admit a decomposition of the form:

\begin{align}
p(abc \vert xyz) = \int \pi(\lambda)  
p(a \vert x \lambda) \mathrm{Tr}(M_{b \vert y} \sigma_{\lambda}^B)  \mathrm{Tr}(M_{c \vert z} \sigma_{\lambda}^C) d\lambda.
\end{align}

That is, the statistics can be explained by a so-called local hidden-state model (LHS), where the local response functions from (\ref{LHV}) of Bob and Charlie are obtained from measurements on a pre-determined quantum state $ \sigma_{\lambda}^B $ respectively $ \sigma_{\lambda}^C $. The set of (unnormalised) conditional states $ \lbrace \sigma^{BC}_{a \vert x} \rbrace $ that Alice can prepare for Bob and Charlie,
the so-called assemblage, is given by

\begin{align}
\sigma^{BC}_{a \vert x} = \mathrm{Tr}_A[(M_{a \vert x} \otimes \mathds{1} \otimes \mathds{1}) \rho],
\end{align}

where $ \mathrm{Tr}_A $ denotes the partial trace and $ \mathrm{Tr}(\sigma_{a \vert x}) = p(a \vert x) $ is the probability that Alice obtains outcome a. Here, the measurement sets of Bob and Charlie $ \lbrace M_{b \vert y} \rbrace $ and $ \lbrace M_{c \vert z} \rbrace  $ are assumed as tomographically complete. Further, note that any LHS can be considered as an LHV, while the converse does not hold \cite{PhysRevA.92.032107}. An assemblage is said to demonstrate steering if it does not admit the decomposition

\begin{align}
\sigma^{BC}_{a \vert x} = \int \pi(\lambda) p(a \vert x \lambda) \rho^{BC}_{\lambda}, 
\end{align}

where $ \rho^{BC}_{\lambda} $ is a separable quantum state shared by Bob and Charlie. \\
\indent We present now the hidden nonlocality scenario in the spirit of \cite{PhysRevLett.111.160402}. In this scenario the parties perform a probabilistic pre-selection according to a desired outcome before the Bell test. Hence, they apply a sequence of measurements on the shared state $ \rho_L $ which can lead to nonlocal correlations even if $ \rho_L $ admits an LHV for POVMs. In particular this idea can be implemented by the use of local filters given by arbitrary Kraus operators $ F_x $, fulfilling $ F_x^{\dagger} F_x \leq \mathds{1}, x \in \lbrace A, B, C \rbrace $ and acting on the respective local Hilbert space of the observers.  
The state which the parties share after filtering is given by

\begin{align}
\rho = \dfrac{F_A \otimes F_B \otimes F_C \ \rho_L \ F_A^{\dagger} \otimes F_B^{\dagger} \otimes F_C^{\dagger} }
{\mathrm{Tr}(F_A \otimes F_B \otimes F_C \ \rho_L \ F_A^{\dagger} \otimes F_B^{\dagger} \otimes F_C^{\dagger} )},  
\end{align}

where the success probability of the filtering is given by the normalization factor.
We say that a state $ \rho_L $ possesses genuine hidden nonlocality if it admits an LHV for POVMs but the state $ \rho $ for some judiciously chosen filters $ F_A, F_B, F_C $ does violate a Bell
inequality. Note that local invertible filters do not change the entanglement character of a given state \cite{RevModPhys.81.865}, i.e. bound entangled states remain bound entangled. Further, by bound entangled states we mean entangled states with positve partial transpose (PPT).   \\
\indent \textit{Methods.}\textemdash In order to derive our results, we will solve two main tasks: We show that the filtered state does violate a Bell inequality and that the state before filtering admits a local
model for POVMs. The first task can be solved efficiently by an iterative sequence of semidefinite programms (SDPs) \cite{Vandenberghe}, using the so-called see-saw \cite{quant-ph/0107093} method. \\
\indent Consider a Bell inequality of the form:

\begin{align}
I = \sum_{a,b,c,x,y,z} c_{abc \vert xyz } \  p(abc \vert xyz) \leq L \label{Bell_ineq},
\end{align}

with given (real) coefficients $ c_{abc \vert xyz } $ and the local bound $ L $. The Bell operator according to this inequality is then given by
\begin{align}
\mathcal{B} = \sum_{a,b,c,x,y,z} c_{abc \vert xyz} \ M_{a\vert x} \otimes M_{b \vert y} \otimes M_{c\vert z}.
\end{align}

The goal is to maximize the quantum value $ Q = \mathrm{Tr}(\mathcal{B} \rho) $ for PPT entangled states $ \rho $. To optimize such an expression over all parties and the state is a problem, which cannot be solved by an SDP in general. However, the see-saw method provides a solution: We fix the measurements for two of the parties for a given state $ \rho $, such that the problem becomes linear in the remaining party, let us say Alice. We maximize the expression $ Q $ subject to the constraints $ M_{a \vert x} \geq 0 $, $ \sum_a M_{a \vert x} = \mathds{1} $, which leads us to the optimal measurements of Alice.
This strategy is iteratively applied over the individual parties and the state, to optimise the quantum value $ Q $, without being guaranteed that it is a global maximum. \\
\indent The second task is more difficult to solve. Even though there exist analytical constructions for LHVs, they mostly restrict to certain classes of states with high symmetry or they are restricted to projective measurements. Recently in \cite{PhysRevLett.117.190401,PhysRevLett.117.190402} a method was presented to construct algorithmically local models, again making use of SDPs. Here, we only point out the main use of this construction (for details see \cite{PhysRevLett.117.190401,PhysRevLett.117.190402}). Consider a discrete set of measurements $ \lbrace M_{a \vert x} \rbrace $ and the target state $ \rho_L $. Further, consider the following SDP: 
\begin{align} \label{SDP}
\text{given} \ \ &\rho_L, \lbrace M_{a \vert x } \rbrace \\ 
\text{find}  \ \ &q^* = \mathrm{max} \ q \nonumber \\
\text{s.t.} \ \ &\mathrm{Tr}_A [( M_{a \vert x} \otimes \mathds{1}) \chi ] = \sum_{\lambda} D_{\lambda} (a \vert x ) \sigma_{\lambda}, \ \ \forall \ a,x \nonumber  \\
&\sigma_{\lambda} \geq 0, \sigma_{\lambda}^{\mathrm{T_B}} \geq 0 \ \forall \ \lambda \nonumber \\
&\eta \chi + (1-\eta) \xi_A \otimes \chi_{BC} = q \rho_L +(1-q) \dfrac{\mathds{1}}{d_A d_B d_C}, \nonumber
\end{align}

where $ \chi $ and $ \sigma_{\lambda} $ are the optimization variables. The SDP can be understood as follows. The first constraint ensures that $ \chi $ does admit an LHS for the finite set of measurements $ \lbrace M_{a \vert x} \rbrace $, where $ D_{\lambda} (a \vert x ) $ are the deterministic strategies of cardinality $ N = (k_A)^{m_A} $, with $ k_A $ outcomes for $ m_A $ settings of Alice's measurements. The local hidden-states $ \sigma_{\lambda} $ have to be separable between Bob and Charlie which is in general a non-trivial task, but for two qubits can simply be enforced by the partial transpose constraint $ \sigma_{\lambda}^{\mathrm{T_B}} \geq 0 $. In the last constraint $ \xi_A $ is an arbitrary density matrix  on Alice's side, $ \chi_{BC} = \mathrm{Tr}_A(\chi) $, and $ \eta $ is the so-called shrinking factor with $ 0 \leq \eta \leq 1 $. The constraint ensures that also a noisy version of the target state $ \rho_L $ admits an LHS, but this time for the continuous set of measurements $ \mathcal{M} $ (e.g. four-outcome POVMs) which was approximated by the discrete set $ \lbrace M_{a \vert x} \rbrace \subset \mathcal{M} $. \\
\indent The SDP is based on the fact that the statistics from noisy measurements on a noiseless state are equal to the statistics of a noisy state with noiseless measurements i.e.,

\begin{align}
\mathrm{Tr}_A [(M^{\eta}_a \otimes \mathds{1}) \chi] = \mathrm{Tr}_A [(M_a \otimes \mathds{1}) \rho_L], \label{LHS_equivalence}
\end{align}

where the target state is defined by

\begin{align}
\rho_L = \eta \chi + (1-\eta)\xi_A \otimes \chi_{BC}, 
\end{align}

and the noisy measurements are given by

\begin{align}
M^{\eta}_a = \eta M_a + (1-\eta) \mathrm{Tr}(\xi_A M_a) \mathds{1},
\end{align}

for any $ M_a \in \mathcal{M} $. \\
Note that because $ \chi $ admits an LHS for the discrete set $ \lbrace M_{a \vert x} \rbrace $, by convexity it admits also a local model for the noisy measurements $ M^{\eta}_a $.
From the equality in (\ref{LHS_equivalence}) it follows that $ \rho_L $ does also admit an LHS for a set of continuous noiseless measurements. \\
\indent Here, the shrinking factor $ \eta $ is the largest number such that all noisy measurements $ M^{\eta}_a $ can be written as a convex mixture of elements from the discrete set $ \lbrace M_{a \vert x} \rbrace $ i.e.,

\begin{align}
M^{\eta}_a  = \sum_x p_x M_{a \vert x }, 
\end{align}
with $ \sum_x p_x = 1 $ and $ p_x \geq 0 \ \forall \ x $. \\
\indent The shrinking factor can only be obtained analytically in the case of qubit projective measurements, but for general measurements it can be obtained by an SDP. For more details see \cite{PhysRevLett.117.190402}.\\
\indent \textit{Results.}\textemdash We now display our main result by first presenting a nonlocal three-qubit bound entangled state and in a second step show that this state originates from local filtering of a different state with an LHS model for POVMs. Consider the (real valued) density matrix in the basis $ \lbrace \vert 000 \rangle, \vert 001 \rangle, \vert 010 \rangle,..., \vert 111 \rangle \rbrace_{ABC} $ given by

\begin{align}
\rho_{NL} = (r_{ij})_{1 \leq i,j \leq 8} \label{rho_NL}
\end{align}

with the following defining entries 

\begin{align}
&r_{11} = 0.0290, r_{12} = r_{13} = r_{15} = -0.0098, \nonumber \\
&r_{14} = r_{16} = r_{17} = r_{23} = r_{25} = r_{35} = -0.0083, \nonumber \\
&r_{18} = r_{27} = r_{36} = r_{45} = 0.0646, \nonumber \\
&r_{22} = r_{33} = r_{55} =  0.0412, \nonumber \\
&r_{24} = r_{26} = r_{34} = r_{37} = r_{56} = r_{57} = -0.0335, \nonumber \\
&r_{28} = r_{38} = r_{46} = r_{47} = r_{58} = r_{67} = -0.0598, \nonumber \\
&r_{44} = r_{66} = r_{77} = 0.1352, \nonumber \\
&r_{48} = r_{68} = r_{78} = 0.0102, r_{88} = 0.4418. \nonumber 
\end{align}

Note that $ \rho_{NL} $ is invariant under partial transpose with respect to any party, as well as invariant under permutation of parties, by construction. Therefore, the state is PPT and also bi-separable with respect to any bipartite cut \cite{PhysRevLett.108.030403,PhysRevA.61.062302}. Nevertheless, using the see-saw method it can be shown to violate number 5 of Sliwa's inequalities \cite{Sliwa2003} (which implies $ \rho_{NL} $ is entangled), which reads 

\begin{align}
I = \langle sym[A_{1} +A_{1}B_{2} - A_{2}B_{2} - A_{1}B_{1}C_{1} \nonumber \\
-A_{2}B_{1}C_{1} + A_2 B_2 C_2] \rangle \leq 3 \label{Silwa5}
\end{align}

where $ sym[X] $ denotes the symmetrization of $ X $ over the three parties, e.g., $ sym[A_1 B_2] = A_1 B_2 + A_1 C_2 + A_2 B_1 + 
A_2 C_1 + B_1 C_2 + B_2 C_1. $ Here, $ A_j = B_j = C_j, j \in \lbrace 1,2 \rbrace $ and $ A_j = M_{1 \vert j} - M_{2 \vert j} $. We choose  \\ $ A_1 = -0.7909 \sigma_z -0.6119 \sigma_x$, $A_2 = -0.2344 \sigma_z + 0.9721 \sigma_x $, which leads to a quantum violation $ Q \approx 3.0152 > 3 $ of inequality (\ref{Silwa5}). Next, we show that $ \rho_{NL} $ can originate from a local state by filtering. Consider the state $ \rho_{L} $ defined via the relation 

\begin{align} \label{rho_L_implicit}
\rho_{NL} = \dfrac{F_A \otimes F_B \otimes F_C \ \rho_{L} \ F_A^{\dagger} \otimes F_B^{\dagger} \otimes F_C^{\dagger} }
{\mathrm{Tr}(F_A \otimes F_B \otimes F_C \ \rho_{L} \ F_A^{\dagger} \otimes F_B^{\dagger} \otimes F_C^{\dagger} )},
\end{align}

with the local filters

\begin{align}
F_A = \begin{bmatrix}
       \phantom{-}0.4310  & -0.2971 \\ -0.2488 & \phantom{-}0.7291
      \end{bmatrix}, \nonumber \\
      F_B = \begin{bmatrix}
       \phantom{-}0.0342 & -0.0808 \\ -0.3664 & \phantom{-}0.8688
      \end{bmatrix}, \nonumber \\
      F_C = \begin{bmatrix}
       \phantom{-}0.3268  & -0.1873 \nonumber \\ -0.1773 & \phantom{-}0.6440
      \end{bmatrix}.
 \end{align}

\begin{figure}[h!]
   \includegraphics[scale = 0.5]{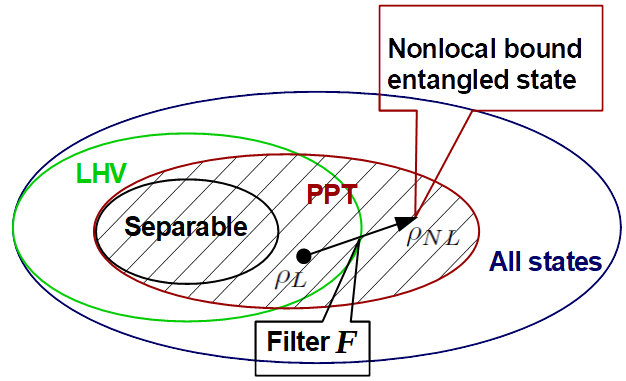} 
   \caption{Schematic overview over the relevant sets of states. The states in the shaded area are undistillable. Our results confirm the existence of bound entangled states with an LHV for POVMs. However, (invertible) local filters $ F $ are able to reveal the hidden nonlocality of these states. They map a state $ \rho_L $ from the set of states admitting an LHV onto a nonlocal state $ \rho_{NL} $.}
   \label{Results_mapping}
 \end{figure}

For more details, see the Supplemental Material \cite{Supplemental_Material}. Note that it is immediately clear that there exists a valid quantum state $ \rho_{L} $ fulfilling the above equation. This can be seen by using the fact that the above local filters are invertible and the only constraint $ F^{\dagger} F \leq \mathds{1} $ can always be achieved, since the filters $ F $ and $ cF $ map onto the same state for any $ c \in \mathbb{C} \setminus \lbrace 0 \rbrace $. \\
\indent In order to finally show that $ \rho_{L} $ does possess genuine hidden nonlocality, we need to show that it admits a local model for all POVMs. Therefore, we use the same parametrization as in \cite{PhysRevLett.117.190402} for Alice's finite set of measurements $ \lbrace M_{a \vert x} \rbrace $. It consists of all relabellings of $ \lbrace P_+, P_-, 0, 0 \rbrace $ where $ P_+ $ is a projector onto the vertex of an icosahedron in the Bloch sphere and $ P_- $ onto the opposite direction, as well as all relabellings of the trivial set $ \lbrace \mathds{1}, 0, 0, 0 \rbrace $. This leads to a set of $ 76 $ elements with a shrinking factor of $ \eta \approx 0.673 $ for $ \xi_A = \mathds{1}/2 $. Note that it is sufficient to consider only extremal POVMs, which for qubits have at most four outcomes \cite{DAriano2005}. The optimization for the LHS, according to (\ref{SDP}) results in $ q^* = 1 $. Hence, $ \rho_{L} $ does admit a local model for POVMs without the need of additional noise. For a graphical illustration of our main results, see Fig. (\ref{Results_mapping}). 

\indent \textit{Conclusions and Outlook.}\textemdash In the present letter, we have shown that a fully bi-separable bound entangled state of three qubits can admit a local model for POVMs, but can give rise to nonlocal correlations when local filters were applied before the Bell test. Hence, we have shown that bound entangled states can possess genuine hidden nonlocality. This is the first example of activation of nonlocality in bound entanglement. Furthermore, this is also the first example for an LHV of a bound entangled state for all POVMs, while previous models were restricted to projective measurements \cite{PhysRevLett.117.190402,PhysRevLett120020506}. One important conclusion of our results is that genuine hidden nonlocality (since it is also possible for bound entangled states) does not imply entanglement distillability. Together with the result of \cite{1367-2630-18-11-113019} it shows that genuine hidden nonlocality and entanglement distillation are inequivalent. Note that since the local model we have constructed is an LHS model, our results are also relevant for the steering scenario.\\ 
\indent It would be interesting to know whether there exist also bound entangled states without hidden nonlocality. Even though we could not prove the existence of such states, we found a bipartite bound entangled state with a local model for POVMs in the so-called filter normal form \cite{Verstraete}, which seems to play an important role for hidden nonlocality. We think therefore, that this state is a good candidate to show bound entanglement without hidden nonlocality. For further details, see the Supplemental Material \cite{Supplemental_Material}. In the future, one should investigate the potential of bound entangled states in the superactivation or even in the asymptotic scenario. Even 20 years after the Peres conjecture, we still learn what bound entangled states are useful for \cite{PhysRevA.99.032334}. In the spirit of these developments it seems to be well-motivated to state an ``inverse Peres conjecture'': all bound entangled states are nonlocal resources in the asymptotic case, see Fig. \ref{Results_sets}.
\newpage

This project has received funding from the European Union's Horizon2020  research  and  innovation  programme  under  the  Marie  Sk\l{}odowska-Curie  grant agreement No 675662 and support from the Federal Ministry of Education and Research (BMBF).

\bibliography{Paper_GHNLPPT.bib}

\bibliographystyle{unsrt}

\appendix*
\section{Supplemental Material for ``Activation of nonlocality in bound entanglement '' }

\indent \textit{Details on the local state $ \rho_L $.}\textemdash In order to give a useful representation of the local state $ \rho_{L} $ from (\ref{rho_L_implicit}) in the main text, one has to understand how to obtain this state. Naturally, there is no hint which states one should investigate in order to try to prove their locality or whether they possess genuine hidden nonlocality.
However, it becomes immediately clear when one inverts the problem and tries to find a local state after we applied local filters on a nonlocal state. Since we choose the filters to be invertible, we can easily find filters which map the local state onto the nonlocal state. The nonlocal state obtained by the see-saw algorithm has by construction a high amount of symmetry, which we decrease by the local filters and then apply the SDP techniques to find an LHS. Afterwards, the inverted filters increase the symmetry of the state again. Therefore, $ \rho_{L} $ is simply given by

\begin{align}
\rho_{L} = \dfrac{G_A \otimes G_B \otimes G_C \ \rho_{NL} \ G_A^{\dagger} \otimes G_B^{\dagger} \otimes G_C^{\dagger} }
{\mathrm{Tr}(G_A \otimes G_B \otimes G_C \ \rho_{NL} \ G_A^{\dagger} \otimes G_B^{\dagger} \otimes G_C^{\dagger} )}, \tag{A1} \nonumber
\end{align}

with the local invertible filters

\begin{align}
G_A = \begin{bmatrix}
       0.7291 & 0.2971 \\ 0.2488 & 0.4310
      \end{bmatrix}, \nonumber \\
      G_B = \begin{bmatrix}
       0.8688  & 0.0808 \\ 0.3664 & 0.0342
      \end{bmatrix}, \nonumber \\
      G_C = \begin{bmatrix}
       0.6440  & 0.1873 \nonumber \\ 0.1773 &  0.3268
      \end{bmatrix}.
 \end{align}

and the nonlocal state $ \rho_{NL} $ defined in Eq. (\ref{rho_NL}) in the main text.

\indent \textit{Local bound entanglement in the filter normal form.}\textemdash
Here, we want to extend our outlook, by presenting a bipartite bound entangled state which admits an LHS for POVMs and is a good candidate to show bound entanglement without hidden nonlocality, as we will argue below.  An important feature of this state is that the state is already in the filter normal form \cite{Verstraete}, which means all single party reduced density matricies are maximally mixed. The filter normal form does play an important role when it comes to hidden nonlocality. For example, the filter normal form does maximize the violation of the CHSH inequality for two-qubits, as well as entanglement monotones \cite{Verstraete}. Further, in \cite{1367-2630-18-11-113019} it was shown that certain Werner states do admit an LHS model, even after arbitrary local filtering. Werner states are also already in the filter normal form. \\
\indent Intuitively, there is no obvious reason why local filters would still be able to activate the nonlocality of such states because they cannot distinguish the \emph{useful} part of a state from white noise. Consider the state, in filter normal form given by

\begin{align}
\sigma = \dfrac{\mathds{1}}{d_A d_B} + \sum\limits_{k=1}^{d_A^2-1} \xi_A H_k^A \otimes H_k^B \tag{A2} \nonumber
\end{align}

\noindent with $ d_A = 2 $, $ d_B = 4 $, the coefficients $ \xi_k $, and the traceless mutually orthonormal matricies $ H_k^A $, $ H_k^B $. Specifically, we choose 
\begin{align}
\xi_1 = \xi_2 = 1.3219, \ \xi_3 = 1.1348, \nonumber 
\end{align}

\noindent and the matricies 
\begin{align}
&H_1^A = \begin{pmatrix}
0 & 0 \\
1 & 0\\
\end{pmatrix},\ H_2^A = \begin{pmatrix}
0 & -1\\
0 & \phantom{-}0\\
\end{pmatrix}, \nonumber 
\\ &H_3^A = \begin{pmatrix}
\dfrac{1}{\sqrt{2}} & \phantom{-}0 \\
0 & -\dfrac{1}{\sqrt{2}}\\
\end{pmatrix}, \nonumber
\end{align}
for Alice's subsystem, as well as

\begin{align}
H_1^B = \begin{pmatrix}
\phantom{-}0 & \phantom{-}0 & \phantom{-}0 & -0.0983 \\
-0.6393 & \phantom{-}0 & \phantom{-}0 & \phantom{-}0 \\
\phantom{-}0 & -0.4158 & \phantom{-}0 & \phantom{-}0 \\
\phantom{-}0 & \phantom{-}0 & -0.6393 & \phantom{-}0 \\
\end{pmatrix}, \nonumber \\
H_2^B = \begin{pmatrix}
\phantom{-}0 & \phantom{-}0.6393 & \phantom{-}0 & \phantom{-}0 \\
\phantom{-}0 & \phantom{-}0 & \phantom{-}0.4158 & \phantom{-}0 \\
\phantom{-}0 & \phantom{-}0 & \phantom{-}0 & \phantom{-}0.6393 \\
\phantom{-}0.0983 & \phantom{-}0 & \phantom{-}0 & \phantom{-}0 \\
\end{pmatrix}, \nonumber \\
H_3^B = \begin{pmatrix}
-0.4859 & \phantom{-}0 & \phantom{-}0 & \phantom{-}0 \\
\phantom{-}0 & -0.5137 & \phantom{-}0 & \phantom{-}0 \\
\phantom{-}0 & \phantom{-}0 & \phantom{-}0.5137 & \phantom{-}0 \\
\phantom{-}0 & \phantom{-}0 & \phantom{-}0 & \phantom{-}0.4859 \\
\end{pmatrix}, \nonumber
\end{align}

for Bob's side. As one can quickly verify, $ \sigma $ is a PPT state.
Nevertheless, it can be shown to be entangled by the SDP techniques presented in \cite{PhysRevA.69.022308}. With the methods described in the main text, we were able to show that $ \sigma $ does admit an LHS model for general POVMs on Alice's side. \\
\indent As argued above, this state is a good candidate to show bound entanglement without hidden nonlocality. However, it is quite complicated to prove our conjecture, due to the fact that many degrees of freedom are involved. If our conjecture turns out to be true, other scenarios like the superactivation or the asymptotic scenario have to be considered. If it turns out that $ \sigma $ can show hidden nonlocality, it would be the first example of a nonlocal bound entangled state in the lowest possible dimension for two parties. So far the smallest dimension for examples of nonlocal bound entangled states is $ 3 \times 3 $ \cite{1405.4502}.  

\end{document}